\documentclass[preprint,pra,aps,superscriptaddress,showpacs,prd]{revtex4}
\usepackage{epsfig,amsmath,graphicx,amssymb}
\def\Om{\Omega}
\def\be{\begin{equation}}
\def\ee{\end{equation}}
\def\bea{\begin{eqnarray}}
\def\eea{\end{eqnarray}}
\def\bes{\begin{subequations}}
\def\ees{\end{subequations}}
\begin{document}
\title{Three-Way Entanglement and Three-Qubit Phase Gate Based on a Coherent
       Six-Level Atomic System}

\author{Chao Hang}
\address{Department of Physics, East China Normal University,
Shanghai 200062, China}

\author{Yun Li}
\address{Department of Physics, East China Normal University,
Shanghai 200062, China}

\author{Lei Ma}
\email{lma@phy.ecnu.edu.cn}
\address{Department of Physics, East China Normal University,
Shanghai 200062, China}

\author{Guoxiang Huang}
\email{gxhuang@phy.ecnu.edu.cn}
\address{Department of Physics, East China Normal University,
Shanghai 200062, China}

\date{\today}


\begin{abstract}

We analyze the nonlinear optical response of a six-level atomic
system under a configuration of electromagnetically induced
transparency. The giant fifth-order nonlinearity generated in such
a system with a relatively large cross-phase modulation effect can
produce efficient three-way entanglement and may be used for
realizing a three-qubit quantum phase gate. We demonstrate that
such phase gate can be transferred to a Toffoli gate, facilitating
practical applications in quantum information and computation.

\end{abstract}

\maketitle


Photons are considered as promising candidates for carrying
quantum information because of their high propagating speed and
negligible decoherence\cite{Nielsen}. Many proposals have come up
for efficiently implementing all-optical quantum information
processing and quantum computation, some of which are based on
linear optics, and others are considered from nonlinear optical
processes. As is well known, Kerr nonlinearity is crucial for
producing photon-photon entanglement and for realizing two-qubit
optical quantum gates. Similarly, higher-order optical
nonlinearities can be used to produce an $N$-way ($N\geq 3$)
entanglement and realize a multi-qubit quantum gate. However,
optical quantum gates can not be efficiently implemented based on
a conventional optical medium. The reason is that either the
optical nonlinearity produced in such medium is very weak, or
there is a very large optical absorption when working near
resonant regime where nonlinear effect is strong.

In recent years, much attention has been paid to the study of
electromagnetically induced transparency (EIT) in resonant atomic
systems\cite{Boller,Harris1997}. By means of the effect of quantum
coherence and interference induced by a control field, the
absorption of a weak probe field tuned to a strong one-photon
resonance can be largely cancelled and hence an initially highly
opaque optical medium becomes transparent.  The wave propagation
in a resonant optical medium with EIT configuration possesses many
striking features. One of them is the significant reduction of the
group velocity of the probe pulse. Another is the giant
enhancement of the Kerr nonlinearity of the optical
medium\cite{lukin,Fleischhauer}. Several suggestions for obtaining
enhanced Kerr nonlinearity and a related large cross-phase
modulation (CPM) by using the EIT effect have been proposed ,
including ``N'' configuration\cite{Schmidt,Kang}, chain-$\Lambda$
configuration\cite{Greentree}, tripod configuration\cite{Petro},
and symmetric six-level configuration\cite{Petrosyan}. Based on
the enhanced Kerr nonlinearity, two-qubit entanglement with
photons and atoms
\cite{Lukin,Paternostro,Payne,Kiffner,Peng,Friedler} has been
investigated and all-optical two-qubit quantum phase gate
(QPG)\cite{Ottaviani,Rebic,Joshi,li} has also been constructed by
using different schemes recently. However, as far as we know, up
to now only a few works\cite{Zubairy} have studied higher-order,
especially the fifth-order optical nonlinearity, and its
applications to multi-photon entanglement and optical phase gates
under practice EIT configurations.

In this work, we investigate the nonlinear optical response and
possible three-way entanglement and three-qubit phase gates based
on a coherent six-level atomic system under an asymmetric EIT
configuration. Our study shows that, due to quantum interference,
fifth-order nonlinearity in such system can be largely enhanced
with a vanishing linear and third-order nonlinear effect. Using
this property the system can be used to produce efficient
three-way entanglement among three optical (probe, signal, and
trigger) fields. We then explore the possibility of employing an
enhanced CPM effect to devise a mechanism of polarization
three-qubit quantum phase gate (QPG). The three-qubit QPG proposed
here is rather robust, and can be easily transferred to a
universal three-qubit Toffoli gate. Although a Toffli gate can be
constructed by other basic quantum gates, its realization in a
more compact way is needed to dramatically reduce the number of
qubit and manipulations that are required to perform a given task.
Although some studies of constructing Toffoli gate with different
systems\cite{Pachos,Duan,Gagnebin,Tame} exist, our work presented
here is for the first time a practical realization of Toffoli gate
in an all-optical way.

\vspace{3mm}

We start with considering a life-time broadened atomic system,
where atoms with six levels (three ground state levels
$|1\rangle$, $|3\rangle$, $|5\rangle$, and three excited state
levels $|2\rangle$, $|4\rangle$, $|6\rangle$) interact with five
laser fields (see Fig.1).
%
\begin{figure}
\centering
\includegraphics[scale=0.5]{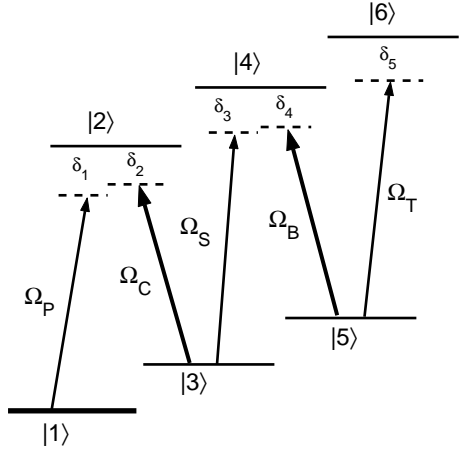}
\caption{\footnotesize The energy-level diagram and excitation
scheme of a life-time broadened six-level atomic system
interacting with two strong, cw control fields of Rabi frequencies
$\Omega_{C}$ and $\Omega_{B}$, and three weak, pulsed (probe,
signal, and trigger) fields of Rabi frequencies $\Omega_{P}$,
$\Omega_{S}$ and $\Omega_{T}$.}
\end{figure}
Such configuration can be realized in Zeeman-splitted alkali atoms
(e.g., the $D1$ line in $^{23}$Na or $^{87}$Rb gas). We assume
that the transitions from $|2\rangle\leftrightarrow|3\rangle$ and
$|4\rangle\leftrightarrow|5\rangle$ are driven  by two strong,
continuous-wave (cw) laser control fields, with Rabi frequencies
$\Om_C$ and $\Om_B$, respectively. The transitions from
$|1\rangle\leftrightarrow|2\rangle$,
$|3\rangle\leftrightarrow|4\rangle$, and
$|5\rangle\leftrightarrow|6\rangle$ are driven by three weak,
pulsed laser fields, called probe field (with Rabi frequency
$\Om_P$) signal field (with Rabi frequency $\Om_S$) and trigger
field (with Rabi frequency $\Om_T$), respectively. Here the Rabi
frequencies associated with the laser fields that drive the atomic
transitions are defined as $\Omega_k=-D_{ij}{\cal E}_k/\hbar$,
where ${\cal E}_k$ denotes the $k$th electric field envelope and
$D_{ij}$ is the relevant electric-dipole matrix element related to
the transition $|i\rangle\leftrightarrow|j\rangle$. The detunings
$\delta_i$ are defined as $\delta_1=(E_2-E_1)/\hbar-\omega_P$,
$\delta_2=(E_2-E_3)/\hbar-\omega_C$,
$\delta_3=(E_4-E_3)/\hbar-\omega_S$,
$\delta_4=(E_4-E_5)/\hbar-\omega_B$, and
$\delta_5=(E_6-E_5)/\hbar-\omega_T$,  where $E_i$ ($i$=1,...,6) is
the energy of the level $|i\rangle$ and $\omega_j$ ($j$=P, C, S,
B, and T) is the frequency of the laser field with the Rabi
frequency $\Omega_j$. The evolution equations for the probability
amplitudes $a_i(t)$ of the atomic state
$|\psi(t)\rangle=\sum_{i=1}^{6}a_{i}(t)|i\rangle$ are
\bes\label{AV}\bea
& &\dot{a}_1=-\frac{\Gamma_1}{2}a_1-i\Omega_P^*a_2,\\
& &\dot{a}_2=-(\frac{\Gamma_2}{2}+i\delta_{1})a_2-i\Omega_Pa_1-i\Omega_Ca_3,\\
& &\dot{a}_3=-(\frac{\Gamma_3}{2}+i\delta_{12})a_3-i\Omega_C^*a_2-i\Omega_S^*a_4,\\
& &\dot{a}_4=-(\frac{\Gamma_4}{2}+i\delta_{13})a_4-i\Omega_Sa_3-i\Omega_Ba_5,\\
& &\dot{a}_5=-(\frac{\Gamma_5}{2}+i\delta_{14})a_5-i\Omega_B^*a_4-i\Omega_T^*a_6,\\
&
&\dot{a}_6=-(\frac{\Gamma_6}{2}+i\delta_{15})a_6-i\Omega_Ta_5,
\eea
\ees
where $\delta_{12}=\delta_1-\delta_2$,
$\delta_{13}=\delta_{12}+\delta_3$,
$\delta_{14}=\delta_{13}-\delta_4$, and
$\delta_{15}=\delta_{14}+\delta_5$. $\Gamma_i$ denotes the decay
rate for the atomic level $|i\rangle$. For the excited state
levels ($|2\rangle$, $|4\rangle$, and $|6\rangle$) these rates
describe the total spontaneous decay rates, while for the ground
state levels ($|1\rangle$, $|3\rangle$, and $|5\rangle$) the
associated decay rates describe dephasing processes.


For solving Eq. (\ref{AV}) we assume that the typical temporal
duration of the probe, signal, and trigger fields is long enough
so that a steady state approximation can be employed. The system's
initial state is assumed to be the ground state $|1\rangle$. When
the intensity of the probe, signal, and trigger is much weaker
than the intensity of both coupling fields, the population in the
ground states $|1\rangle$ is not depleted even when the system
reaches the steady state, i.e. $a_0\approx1$. We solve Eq.
(\ref{AV}) under these consideration and obtain the following
expressions for the susceptibilities of three weak fields
\bes\label{CHI1}
\bea
& &\chi_P\simeq
\chi_{P}^{(1)}+\chi_{PS}^{(3)}|E_S|^2+\chi_{PT}^{(3)}|E_T|^2+\chi_{PST}^{(5)}|E_S|^2|E_T|^2,\\
& &\chi_S\simeq
\chi_{SP}^{(3)}|E_P|^2+\chi_{SPT}^{(5)}|E_P|^2|E_T|^2,\\
& &\chi_T\simeq\chi_{TPS}^{(5)}|E_P|^2|E_S|^2,
\eea
\ees
with
\bes\label{CHI2}
\bea
\chi_{P}^{(1)}&=&\frac{N_a|D_{12}|^2}{\hbar\epsilon_0}\frac{d_3}{d_2d_3-|\Omega_C|^2},\\
\chi_{PS}^{(3)}&=&-\frac{N_a|D_{12}|^2|D_{34}|^2}{\hbar^3\epsilon_0}\frac{d_5}
{(d_4d_5-|\Omega_B|^2)(d_2d_3-|\Omega_C|^2)},\\
\chi_{PT}^{(3)}&=&-\frac{N_a|D_{12}|^2|D_{56}|^2}{\hbar^3\epsilon_0}\frac{d_3d_4}
{d_6(d_4d_5-|\Omega_B|^2)(d_2d_3-|\Omega_C|^2)},\\
\chi_{PST}^{(5)}&=&\frac{N_a|D_{12}|^2|D_{34}|^2|D_{56}|^2}{\hbar^5\epsilon_0}\frac{1}
{d_6(d_4d_5-|\Omega_B|^2)(d_2d_3-|\Omega_C|^2)},\\
\chi_{SP}^{(3)}&=&\frac{N_a|D_{12}|^2|D_{34}|^2}{\hbar^3\epsilon_0}\frac{d_5|\Omega_C|^2}
{(d_4d_5-|\Omega_B|^2)|d_2d_3-|\Omega_C|^2|^2},\\
\chi_{SPT}^{(5)}&=&-\frac{N_a|D_{12}|^2|D_{34}|^2|D_{56}|^2}{\hbar^5\epsilon_0}\left[\frac{|\Omega_C|^2}
{d_6(d_4d_5-|\Omega_B|^2)|d_2d_3-|\Omega_C|^2|^2} \right. \nonumber\\
& &\left. +\frac{d_4^*d_5|\Omega_C|^2}{d_6^*|d_4d_5-|\Omega_B|^2|^2|d_2d_3-|\Omega_C|^2|^2}\right],\\
\chi_{TPS}^{(5)}&=&\frac{N_a|D_{12}|^2|D_{34}|^2|D_{56}|^2}{\hbar^5\epsilon_0}\frac{|\Omega_B|^2|\Omega_C|^2}
{d_6|d_4d_5-|\Omega_B|^2|^2|d_2d_3-|\Omega_C|^2|^2}.
\eea
\ees
Here $\chi^{(1)}$, $\chi^{(3)}$, and $\chi^{(5)}$ denote the
linear, third-order, and fifth-order susceptibilities
corresponding each field, star denotes the complex conjugation,
and $N_a$ is the density of the atomic gas.  We have defined
$d_2=\delta_1-i\Gamma_2/2$ ,$d_3=\delta_{12}-i\Gamma_3/2$,
$d_4=\delta_{13}-i\Gamma_4/2$, $d_5=\delta_{14}-i\Gamma_5/2$, and
$d_6=\delta_{15}-i\Gamma_6/2$.

Above results show that the nonlinear susceptibilities associated
with CPM can be largely enhanced. This can be seen from Eq.
(\ref{CHI2}) that, under the conditions $d_{3}\approx
d_{5}\approx0$\cite{note1}, the fifth-order susceptibilities
remain and have comparably giant values while the linear and
third-order susceptibilities being efficiently suppressed. Thus
under such conditions the system provides only a fifth-order
nonlinear effect. In addition, the imaginary parts of the linear
and nonlinear susceptibilities given above are much smaller than
their relevant real parts under the (EIT) condition
$|\Omega_{P}|^2$, $|\Omega_{S}|^2$, and $|\Omega_{T}|^2\ll
|\Omega_{C}|^2, |\Omega_{B}|^2$, which results in quantum
interferences between the states
$|1\rangle\leftrightarrow|3\rangle$ and
$|3\rangle\leftrightarrow|5\rangle$, making the population in the
excited states be small thus very low absorption for the probe,
signal, and trigger fields.

The results (\ref{CHI1}) and (\ref{CHI2}) enable one to asses the
group velocities of the probe, signal, and trigger fields. As we
know, group velocities have to be comparable and small in order to
achieve an effective CPM effect\cite{Luk}. Unlike the six-level
scheme studied in\cite{Petrosyan}, the present scheme is not
symmetric and hence probe, signal, and trigger group velocities
are generally not equal. Assuming working at the center of the
transparency window for the probe and signal fields, i. e.
$\delta_{12}\approx\delta_{14}\approx0$, and neglecting the
dephasing rates $\Gamma_3$ and $\Gamma_5$, which are typically
much smaller than all the other parameters, we obtain the
expressions of group velocities from (\ref{CHI1})-(\ref{CHI2}) for
the the probe, signal, and trigger fields as
\bes \label{GRO}
\bea
v_{g}^{P}&\simeq&\frac{2\hbar\epsilon_0c|\Omega_{C}|^2|\Omega_{B}|^2}
{N_a|D_{12}|^2\omega_P(|\Omega_{B}|^2+|\Omega_{S}|^2+|\Omega_{T}|^2\beta_1
-|\Omega_{S}|^2|\Omega_{T}|^2\beta_2)},\\
v_{g}^{S}&\simeq&\frac{2\hbar\epsilon_0c|\Omega_{C}|^2|\Omega_{B}|^2}
{N_a|D_{34}|^2\omega_S|\Omega_{P}|^2(1+|\Omega_{T}|^2\beta)},\\
v_{g}^{T}&\simeq&\frac{2\hbar\epsilon_0c|\Omega_{C}|^2|\Omega_{B}|^2}
{N_a|D_{56}|^2\omega_T|\Omega_{P}|^2|\Omega_{S}|^2\beta},
\eea
\ees
with
$\beta_1=(\delta_3\delta_5+\Gamma^2/4)/(\delta_5^2+\Gamma^2/4)$,
$\beta_2=[(\delta_3\delta_5+\Gamma^2/4)(\delta_5^2+\Gamma^2/4)/|\Omega_B|^2+
(\delta_1\delta_5+\Gamma^2/4)(\delta_5^2+\Gamma^2/4)/|\Omega_C|^2-(\delta_5^2-\Gamma^2/4)]
/(\delta_5^2+\Gamma^2/4)^2$, and
$\beta=(\delta_5^2-\Gamma^2/4)/(\delta_5^2+\Gamma^2/4)^2$. For
simplicity for getting above results we have set
$\Gamma_2=\Gamma_4=\Gamma_6=\Gamma$. We note that three velocities
$v_g^P$, $v_g^S$, and $v_g^T$ can be made both small and equal by
properly adjusting the Rabi frequencies and detunings (see the
example given below).

\vspace{3mm}

Significant three-body interaction is a key ingredient for the
production of three-way entanglement and construction of
three-qubit QPG. In our system, such interaction is realized by
the giant CPM effect, in which an optical field acquires a large
phase shift conditional to the state of the other two optical
fields. A three-qubit QPG can be represented by the input-output
relations $|\alpha\rangle_{P}|\beta\rangle_{S}|\gamma\rangle_{T}
\rightarrow
\exp(i\phi_{\alpha\beta\gamma})|\alpha\rangle_{P}|\beta\rangle_{S}|\gamma\rangle_{T}$
where $\alpha,\beta,\gamma=0,1$ denote three-qubit basis.

We choose two orthogonal light polarizations $|\sigma^{-}\rangle$
and $|\sigma^{+}\rangle$ to encode binary information for each
qubit. We assume the six-level system shown in Fig. 1 is
implemented only when the probe, signal, and trigger all have
$\sigma^{+}$ polarization. For a $\sigma^{-}$ polarized probe
there is no sufficiently close excited state to which level
$|1\rangle$ couples and no population in $|3\rangle$ and
$|5\rangle$ to drive the signal and trigger transitions. So the
probe, signal, and trigger only acquire the trivial vacuum phase
shift $\phi_{0}^i=k_iL$ (i=P, S, T; L denotes the length of the
medium). When the probe and signal are $\sigma^{+}$ and
$\sigma^{-}$ polarized, the probe, subject to the EIT produced by
the $|1\rangle-|2\rangle-|3\rangle$ levels $\Lambda$
configuration, acquires a linear phase shift
$\phi_\Lambda^P=k_PL(1+2\pi\chi_P^{(1)})$, while the signal and
trigger acquire again the vacuum shifts $\phi_{0}^S$ and
$\phi_{0}^T$. For a $\sigma^{+}$, $\sigma^{+}$ and $\sigma^{-}$
polarized probe, signal and trigger, the first two fields will
acquire nonlinear cross-phase shifts $\phi_{3-order}^P$ and
$\phi_{3-order}^T$ containing a third-order nonlinear effect,
while the last acquire still the vacuum shift $\phi_{0}^T$. Only
when all three pulses have the ``right'' polarization, they
acquire nonlinear cross-phase shifts $\phi_{5-order}^P$,
$\phi_{5-order}^S$ and $\phi_{5-order}^T$ containing both three-
and fifth-order nonlinear effects.

Assuming that the input probe, signal, and trigger polarized
single photon wave packets can be expressed as a superposition of
the circularly polarized states\cite{Ottaviani,Rebic,Joshi}, i.e.
$|\psi_i\rangle=1/\sqrt{2}|\sigma^-\rangle_i+1/\sqrt{2}
|\sigma^+\rangle_i$ ($i=P, S, T$), where
$|\sigma^\pm\rangle_i=\int d\omega \xi_i(\omega)
a_\pm^\dagger(\omega)|0\rangle$ with $\xi_i(\omega)$ being a
Gaussian frequency distribution of incident wave packets centered
at frequency $\omega_i$. The photon field operators undergo a
transformation while propagating through the atomic medium of
length $L$, i.e. $a_\pm(\omega)\rightarrow
a_\pm(\omega)\exp\{i\omega/c\int_{0}^{L}dz n_\pm(\omega,z)\}$.
Assuming that $n_\pm(\omega,z)$ (the real part of the refractive
index) varies slowly over the bandwidth of the wave packet
centered at $\omega_i$, one gets
$|\sigma^\pm\rangle_i\rightarrow\exp{(-i\phi_\pm^i)}|\sigma^\pm\rangle_i$,
with $\phi_\pm^i=\omega/c\int_{0}^{L}dz n_\pm(\omega_i,z)$. Thus,
the truth table for a polarization three-qubit QPG using our
configuration reads:
\bes\label{Tru}\bea & & |\sigma^-\rangle_P|\sigma^\pm\rangle_S|
\sigma^\pm\rangle_T\rightarrow\exp{[-i(\phi_0^P+\phi_0^S+\phi_0^T)]}
|\sigma^-\rangle_P|\sigma^\pm\rangle_S|\sigma^\pm\rangle_T,\\
& &
|\sigma^+\rangle_P|\sigma^-\rangle_S|
\sigma^\pm\rangle_T\rightarrow\exp{[-i(\phi_\Lambda^P+\phi_0^S+\phi_0^T)]}
|\sigma^+\rangle_P|\sigma^-\rangle_S|\sigma^\pm\rangle_T,\\
& &
|\sigma^+\rangle_P|\sigma^+\rangle_S|
\sigma^-\rangle_T\rightarrow\exp{[-i(\phi_{3-order}^P+\phi_{3-order}^S+\phi_0^T)]}
|\sigma^+\rangle_P|\sigma^+\rangle_S|\sigma^-\rangle_T,\\
& &
|\sigma^+\rangle_P|\sigma^+\rangle_S|\sigma^+
\rangle_T\rightarrow\exp{[-i(\phi_{5-order}^P+\phi_{5-order}^S+\phi_{5-order}^T)]}
|\sigma^+\rangle_P|\sigma^+\rangle_S|\sigma^+\rangle_T.
\eea\ees
with $\phi_{3-order}^P=\phi_\Lambda^P+\phi_{PS}^P$,
$\phi_{3-order}^S=\phi_0^S+\phi_{SP}^S$,
$\phi_{5-order}^P=\phi_\Lambda^P+\phi_{PS}^P+\phi_{PT}^P+\phi_{PST}^P$,
$\phi_{5-order}^S=\phi_0^S+\phi_{ST}^S+\phi_{SPT}^S$, and
$\phi_{5-order}^T=\phi_0^T+\phi_{TPS}^T$. Explicitly, they are
given by
\bes
\bea & & \phi_{PS}^P=k_PL\frac{\pi^{3/2}\hbar^2|
\Omega_S|^2}{4|D_{34}|^2}{\rm Re}[\chi_{PS}^{(3)}]
\frac{\text{erf}(\xi_{PS})}{\xi_{PS}},\\
& & \phi_{PT}^P=k_PL\frac{\pi^{3/2}\hbar^2|
\Omega_T|^2}{4|D_{56}|^2}{\rm Re}[\chi_{PT}^{(3)}]
\frac{\text{erf}(\xi_{PT})}{\xi_{PT}},\\
& & \phi_{PST}^P=k_PL\frac{\pi^{3/2}\hbar^4|
\Omega_S|^2|\Omega_T|^2}{4|D_{34}|^2|D_{56}|^2}
{\rm Re}[\chi_{PST}^{(5)}]\frac{\text{erf}(\xi_{PST})}{\xi_{PST}},\\
& & \phi_{ST}^S=k_SL\frac{\pi^{3/2}\hbar^2|
\Omega_T|^2}{4|D_{56}|^2}{\rm Re}[\chi_{ST}^{(3)}]\frac{\text{erf}(\xi_{ST})}{\xi_{ST}},\\
& & \phi_{SPT}^S=k_SL\frac{\pi^{3/2}\hbar^4|
\Omega_P|^2|\Omega_T|^2}{4|D_{12}|^2|D_{56} |^2}{\rm
Re}[\chi_{SPT}^{(5)}]\frac{\text{erf}
(\xi_{SPT})}{\xi_{SPT}},\\
& & \phi_{TPS}^T=k_TL\frac{\pi^{3/2}\hbar^4|
\Omega_P|^2|\Omega_S|^2}{4|D_{12}|^2|D_{34}|^2} {\rm
Re}[\chi_{TPS}^{(5)}]\frac{\text{erf}(\xi_{TPS})}{\xi_{TPS}}. \eea
\ees
where $\xi_{Pi}=\sqrt{2}L(1-v_g^P/v_g^i)/(\tau_{i}v_g^P)$ (i=S, T)
and
$\xi_{ijk}=\sqrt{2}L[(1-v_g^i/v_g^j)^2/\tau_{j}^2v_g^{i2}+(1-v_g^i/v_g^k)^2/\tau_{k}^2v_g^{i2}]^{1/2}$
(i, j, k=S, P, T) with $\tau_{i}$ being the time duration of the
pulse. If the group velocity matching is satisfied, i.e.
$\xi\rightarrow 0$, the $\text{erf}[\xi]/\xi$ reaches the maximum
value $2/\sqrt{\pi}$.

A three-way entanglement can be calculated by  ``residual
entanglement'', which indicates the amount of entanglement among
the probe, signal and trigger that cannot be accounted for by the
entanglements of arbitrary two weak fields. As in Ref.
\cite{Wootters}, the residual entanglement for a three-qubit pure
state can be written as follows:
\be \zeta_{PST}={\cal C}^2_{P(ST)}-{\cal C}^2_{PS}-{\cal
C}^2_{PT}=2(\lambda_1^{PS}\lambda_2^{PS}+\lambda_1^{PT}\lambda_2^{PT}),
\ee
where $\lambda_1^{PS}$ and $\lambda_2^{PS}$ are respectively the
square roots of two eigenvalues of $\rho_{PS}\tilde{\rho}_{PS}$,
$\lambda_1^{PT}$ and $\lambda_2^{PT}$ are also defined in a
similar way. The reduced density matrix
$\rho_{PS}=\text{Tr}_T(\rho_{PST})$ with $\rho_{PST}$ being the
density matrix of the output state, and
$\tilde{\rho}_{PS}=\sigma^P_y\otimes\sigma^S_y\rho_{PS}^*\sigma^P_y\otimes\sigma^S_y$
with $\sigma_y$ being the $y$-component of the Pauli matrix.

We now consider a practical system working with ultra-cold
$^{87}$Rb atomic gas, in which Doppler effect is made small. Atoms
are confined in a magneto-optical trap, where the pertinent lower
and upper levels are 5$S_{1/2}$, $F_L=1$ , and 5$P_{1/2}$,
$F_U=2$. The Zeeman shift of the sublevels in the lower and upper
level can be adjusted by the intensity of an applied magnetic
field. We take
$\delta_{1}=\delta_{2}=40.0\times10^{7}\,\,\text{s}^{-1}$,
$\delta_{3}=\delta_{4}=-40.0\times10^{7}\,\,\text{s}^{-1}$
($\delta_{12}=\delta_{14}=0$),
$\delta_{5}=2.5\times10^{7}\,\,\text{s}^{-1}$
($\delta_{5}\gg\Gamma/2$ should be satisfied to ensure a small
absorption), $\Omega_{C}=8.0\times10^{7}\,\,\text{s}^{-1}$,
$\Omega_{B}=5.2\times10^{7}\,\,\text{s}^{-1}$,
$\Omega_{P}=2.4\times10^{7}\,\,\text{s}^{-1}$,
$\Omega_{S}=2.5\times10^{7}\,\,\text{s}^{-1}$,
$\Omega_{T}=1.4\times10^{7}\,\,\text{s}^{-1}$,
$\Gamma=0.5\times10^7\,\,\text{s}^{-1}$, and
$N_a=10^{12}\,\,\text{cm}^{-3}$. The probe, signal and trigger
have a mean amplitude of about one photon when the beams are
tightly focused and has a time duration about one microsecond.
With the given parameters, one recognize that the system remains
only the fifth-order susceptibilities and acquire the nontrivial
nonlinear phase shifts entirely caused by the fifth-order
nonlinearity only when all weak fields have the ``right''
polarization. Based on which, the pase-gating mechanism is
presented. The group velocities of the weak fields read
$v_g^P\simeq5.5\times10^{3}$\,\,m/s,
$v_g^T\simeq6.0\times10^{3}$\,\,m/s, and
$v_g^T\simeq8.1\times10^{3}$\,\,m/s. A total nonlinear phase shift
of $5\pi$ radians can be obtained for $L\simeq0.095$cm, and the
residual entanglement $\zeta_{PST}\simeq25\%$. The imaginary part
of the fifth-order susceptibilities is one order of magnitude
smaller than the real part, and hence be neglected safely.

With the above parameters, we realize an operation
$\hat{U}=|000\rangle\langle000|+|001\rangle\langle001|+|010\rangle\langle010|+|011\rangle\langle011|
+|100\rangle\langle100|+|101\rangle\langle101|+|110\rangle\langle110|-|111\rangle\langle111|$.
By applying a single qubit rotation $\hat{R_i}$ to the trigger
field where \be
\hat{R}_i(\theta,\varphi)=\begin{pmatrix}\cos{\dfrac{\theta}{2}}&
ie^{-i\varphi}\sin{\dfrac{\theta}{2}}\\-ie^{i\varphi}
\sin{\dfrac{\theta}{2}}&-\cos{\dfrac{\theta}{2}}\end{pmatrix},\ee
we can easily obtain the Toffoli gate by
$\hat{U}_{\text{Toffoli}}=\hat{R}_T(\pi/2,\pi,2) \hat{U}
\hat{R}^{-1}_{T}(\pi/2,\pi,2)$. The explicit operation is
illustrated in Fig. 2.
\begin{figure}
\centering
\includegraphics[scale=0.5]{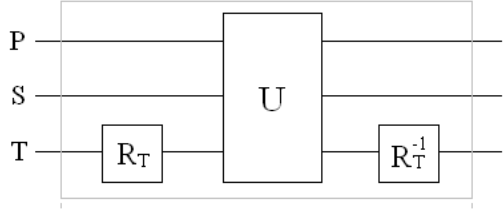}
\caption{\footnotesize The quantum circuit for realizing the
Toffoli gate.}
\end{figure}

\vspace{3mm}


To sum up, we have investigated the nonlinear optical response of
a six-level atomic system under a configuration of
electromagnetically induced transparency. The resultant giant
fifth-order nonlinearity and vanishing linearity and third-order
nonlinearity provided by the system can produce efficient
three-way entanglement among the weak probe, signal, and trigger
laser pulses. Unlike \cite{Zubairy}, here we have addressed a
feasible method to satisfy the group velocity matching among three
optical pulses without using isotopes or solid quantum dots. In
addition, we have studied the possibility of implementing a robust
three-qubit QPG, which can be further transferred to a Toffoli
gate by applying a single qubit rotation. The practical
realization of such a six-level system is easily achievable in a
alkali atomic system in a gas cell. The results provided in this
work may be useful for guiding experimental realization of
three-way entanglement and three-qubit phase gates and
facilitating practical applications in quantum information and
computation.

\acknowledgments The work was supported by the Key Development
Program for Basic Research of China under Grant Nos. 2001CB309300
and 2005CB724508, and NSF-China under Grant Nos. 10434060 and
90403008.



\begin{references}
\bibitem{note1}Since the decay rates in the ground states $|j\rangle$ ($j=1,3,5$)
are small,  $d_3\approx$ 0 and $d_5\approx$ 0 are the the
conditions of EIT-like two- and four-photon resonances,
respectively.
\end{references}
\end{document}